\documentstyle[11pt,newpasp,twoside,epsf]{article}
\def\lapp{\ifmmode\stackrel{<}{_{\sim}}\else$\stackrel{<}{_{\sim}}$\fi}
\def\gapp{\ifmmode\stackrel{>}{_{\sim}}\else$\stackrel{>}{_{\sim}}$\fi}

\reversemarginpar

\begin{document}
\title{IAU 177 --- A week in review}
\author{D.R.~Lorimer}
\affil{Arecibo Observatory, HC3 Box 53995, Arecibo, PR 00612, USA}

\section{Introductory Remarks}

The onerous task of summarising meetings such as these falls
traditionally on some poor unsuspecting soul. Having accepted this
duty (and I still can't believe I ever agreed to it!) it's now time to
look back on what has been one of the most enjoyable meetings I've had
the good fortune to attend.

As always on these occasions, the conference summariser is given a
free rein to review the meeting as he/she feels appropriate. As we all
know, this often results in a significant deviation from the ``ideal
conference summary'' --- an unbiased account of the proceedings. This
review is no exception and, as is the tradition, I shall make it very
clear from the start that the following paragraphs only scratch the
surface of the many varied topics discussed at this meeting.  I choose
to concentrate on the topics I was most interested in, and the new
results that really caught my attention. As such, the many excellent
presentations on pulsar emission theory are therefore not mentioned
here (see however the preceding review by Melrose).  My apologies to
all such participants who do not get a mention in this review. This
merely reflects my personal bias as an observer who takes the
``black-box'' approach to pulsars and life in general.

In all we listened to 95 contributed talks and pondered over 180
posters. Topics covered included pulsar surveys, the interstellar
medium, anomalous X-ray pulsars, magnetars, single pulse studies,
radio emission phenomenology, pulsar timing, supernova remnants,
interactions with companion stars, high energy observations, neutron
star emission theories, fundamental astrometry, general relativity,
neutron star demography, plasma physics and the future of the
field. Many of us also heard the excellent public lecture given by
Jocelyn Bell-Burnell (a {\sl tour de force} presentation of astronomy
to the general public) as well as additional evening sessions
discussing orthogonal polarisation modes and magnetars.  There is
clearly much to talk about and, without further ado, I will begin.

\section{Meeting Highlights}

\subsection{Pulsar Surveys}

The week got off to a fine start with talks by Camilo and Manchester
on the latest results from the Parkes Multibeam survey. With
13 $\lambda$ 21-cm 25 K receivers on the sky, along with $13 \times 2
\times 288$-MHz filterbanks, the system is presently making major
contributions in a number of different pulsar search projects. In its
main use for the Galactic plane survey described by Camilo, the system
achieves a sensitivity of 0.15 mJy in 35 min and covers about
one square degree of sky per hour of observing --- far beyond
the capabilities of any other system at present.

The staggering present total of 439 new pulsars from an analysis of
just under half the total data leads the team to predict that the
final body count should be over 600. Such a large haul is resulting in
significant numbers of interesting individual objects: Several of the
new pulsars are observed to be spinning down at high rates, suggesting
that they are young objects with large magnetic fields. The inferred
age for the 400-ms pulsar J1119$-$6127, for example, is only 1.6
kyr. Another member of this group is the 4-s pulsar J1814$-$1744, an
object that may fuel the ever-present ``injection'' controversy
surrounding the initial spin periods of neutron stars. Further studies
of these objects to look for extended radio/optical emission (i.e.~supernova
remnants/bow shocks) will undoubtably help us to understand the birth
properties of pulsars.

A number of the new discoveries from the survey have orbiting
companions.  Several low-eccentricity systems are known where the
likely companion is a white dwarf star. Two probable double neutron
star systems are presently known: J1811$-$1736 is in an 18-d highly
eccentric orbit, while J1141$-$65 has a lower eccentricity but with an
orbital period of only 4.75 hr. The fact that J1141$-$65 may have a
characteristic age of just over 1 Myr implies that the likely
birth-rate of such objects may be large. Whilst it is premature to
start speculating on statistics of one object, it is clear that these
binary systems and the many which will undoubtably come from this
survey will teach us a lot about the still poorly-understood
population of double neutron star systems.

The most massive binary system from the multibeam survey to date is
J1740$-$3052, whose orbiting companion must be at least 11
M$_{\odot}$. As Manchester mentioned, recent optical observations
reveal a K-supergiant as being the likely companion star in this
system. With such high-mass systems in the Galaxy, not to mention the
massive X-ray binary systems, surely it is only a matter of time
before a radio pulsar will be found orbiting a stellar-mass black
hole. Future searches for these elusive beasts in globular clusters,
where long integration times are common, would do well to utilise
Ransom's novel technique to detect short orbital period binary systems.

We also heard during the meeting that the Parkes multibeam system has
not only been finding young and distant pulsars along the Galactic
plane.  Two other search projects were described in the posters by
Edwards et al.~and Freire et al. Edwards et al.~have been using all 13
beams to do a ``quick'' (5 min per pointing) search at intermediate
Galactic latitudes ($5^{\circ} \leq |b| \leq 15^{\circ}$). The
discoveries of 8 ``recycled'' pulsars during this search, not to
mention 50 long-period objects, strongly support a recent suggestion
by Toscano et al.~that an L-band search of intermediate latitudes is
an excellent means of finding millisecond pulsars which, as they and
others demonstrated (see Kramer's review) now appear to have
significantly flatter spectra than previously thought.

Freire et al., on the other hand, have used only the central beam to
search for new pulsars in the globular cluster 47~Tucanae. This has
revealed no less than 10 new binary millisecond pulsars, bringing the
total number of pulsars in this cluster to 21! The new discoveries
include a 95-min binary system. This is presently the shortest
orbital period for a radio pulsar binary. Clearly these are exciting
times for the pulsar hunting community and the coming months will
undoubtably throw up further surprises. 

\subsection{Gravitational Wave Astronomy}

While pulsar searches in the radio and X-ray regimes are enjoying a
renaissance due to new instrumentation and advances in sensitivity,
our colleagues in the gravitational wave community are about to open a
whole new window on the Universe. As Schutz reported during his talk,
a number of sensitive detectors are about to come on-line. Specifically, 
these are the two LIGO sites in the US, the GEO detector in Hannover
(a joint German/UK project) and, later, the Italian VIRGO
project. Possible events that are being targeted by these detectors
include: (1) compact object mergers at cosmological distances; (2)
continuous emission from spinning neutron stars (though detecting even
nearby pulsars will challenge the limits of present sensitivity); (3)
stellar collapse in the Galaxy.
Millisecond pulsar timing still remains an attractive means of probing
the long-wavelength regime, with space interferometers
still some way off.

In anticipation of the data that will soon be pouring off these
detectors, astronomers at various institutes around the world have
been putting a tremendous effort into understanding the numerous
signal processing/detection hurdles they face. As always, the main
problem is one of sensitivity... astronomers are trying to measure
events that produce a detector ``strain'' of order $10^{-23}$ or less!
In addition, as we heard, analyses of $10^7$ s time series are
required with the initial detector systems. The requirement to deal with
time series of this length is prompting novel techniques to enable
even present state-of-the-art computers analyse the data within a
Hubble time. Radio and X-ray pulsar hunters would do well to stay
in touch with this exciting field for new tricks in the future.

\subsection{Radio-Quiet Neutron Stars and Supernova Remnants}

The so-called ``radio-quiet neutron stars'' were the subject of a
number of presentations during this meeting. This term includes the
soft gamma-ray repeaters (SGRs; thought to be magnetars) and the
anomalous X-ray pulsars (AXPs) and the enigmatic Geminga pulsar (see
below). One significant advance in the high energy observations has
been in the timing analyses that are now possible. The new Rossi X-ray
phase-coherent timing observations of two AXPs by Kaspi et al.~provide
a nice confirmation that these neutron stars are spinning down in a
similar manner to most radio pulsars. Whilst there seems to be
overwhelming evidence that the AXPs are young neutron stars at the
centres of supernova remnants (see e.g.~the contributions by Gotthelf
and Gaensler) I certainly got the impression during some of the
discussions that whilst the evidence associating SGRs with supernova
remnants is tantalising, it is presently only circumstantial.

Part of the problem in making the link arises because of uncertainties
in independently measuring the distances to the remnants and the
SGRs. An interesting development in this regard is the detection of
dispersed low-frequency radio pulses from SGR 1900+14 by Shitov
and collaborators. The dispersion measure obtained implies a distance
of 6 kpc for the most recent electron density model.  Future combined
radio and high-energy timing analyses to monitor the spin-down of this
magnetar will be most interesting, particularly if e.g.~regular radio
observations reveal magnetar glitches. It presently remains a mystery
why this pulsar and Geminga are not detected at higher radio frequencies 
e.g.~430 MHz, this is despite numerous observing campaigns at Arecibo, 
Jodrell Bank and the VLA (see Kassim \& Lazio's poster and references
therein).

\subsection{Pulsar Timing}

A number of recent and interesting pulsar timing results were
presented during the week. High-precision timing of millisecond
pulsars continues to demand careful attention to understanding the
properties of the telescope and data taking system in order to reap
the wealth of astrophysical information that these clocks have to tell
us. This was well highlighted during Britton's study of the systematic
effects of polarisation on the Parkes timing model residuals obtained
for the bright, nearby millisecond pulsar J0437$-$4715. The clever use
of the ``invariant profile'' (Stokes $I^2-Q^2-U^2-V^2$, as opposed to
the total intensity) in the timing analysis is an elegant means (for
weakly polarised pulsars) to circumvent the many non-trivial steps in
a proper polarisation calibration analysis. The quality of the
residuals obtained via this relatively simple analysis should inspire
other observers to try this technique as a means of improving their
timing precision.

Recent results from timing observations that have (finally!) resumed
after the Arecibo upgrade include further confirmation of the orbital
decay of the binary pulsar B1913+16 predicted by general relativity
(see Taylor's paper), as well as a continuing study of the pulse
profile evolution of this pulsar caused by geodetic precession of the
orbit (see Weisberg's paper). Post-upgrade
Arecibo observations of the ``planets pulsar'' B1257+12, when combined
with Effelsberg and pre-upgrade Arecibo observations, and a clever
perturbation analysis, have permitted the orbital inclination of two
of the planets and hence their absolute masses to be determined (see
papers by Wolsczcan and Konacki).

On the subject of planets, a long-awaited update on the timing of the
pulsar B1828$-$11 was presented by Lyne. The curious behaviour of this
system was first reported by Bailes et al.~in the proceedings of the
``Planets around Pulsars'' meeting back in 1992 where the timing
residuals showed a strong periodicity indicative of a sum of three
sinusoids that could be attributed to the Doppler shifting of the
pulsar period by orbiting planetary bodies. The new results presented
here show a strong correlation between pulse profile changes and the
periodic oscillations seen in the timing residuals. Whilst this would
rule out a planetary origin, an interesting twist in the story is
that the phases of the sinusoids sum to a constant value --- as seen
in e.g.~the Jovian satellites. Presently, the most plausible explanation
seems to be free precession of the neutron star. If correct, this
would be the first clear detection of such an effect. It remains a
mystery why similar behaviour is not seen in other young pulsars.

The very useful contributions that can be made by smaller radio
telescopes, which are being used to perform intense observing
campaigns on selected objects, was highlighted several times during
the week. Wolszczan and collaborators have now been performing regular
timing observations with the 32-m telescope in Torun for several
years. Their growing database of observations is being used to search
for planets around pulsars, and to investigate pulsar scintillation
(see papers by Lewandowski et al). The Crab pulsar still continues to
surprise astronomers after almost 30 years of regular monitoring with
the remarkable echoing events (ghost pulse profile components)
seen in observations made using the 85-foot telescope at Green Bank by
Backer and the 42-foot telescope at Jodrell Bank (Smith \&
Lyne). Current interpretations of the echos, which have now been seen
on several occasions, are that they are caused by ionised shells
drifting around the nebula, or intrinsic to the pulsar magnetosphere.

\subsection{Neutron Star Demography}

Significant progress in our understanding of the Galactic population
of neutron stars is being made, and Cordes' presentation summarised
the present status of an ambitious attempt his group is making to
model the population. A big improvement that this model has over
previous studies is the use of a self-consistent beaming model.
Assuming that the core-cone model of the radio beam is correct, and
Desphande and Rankin's polar cap maps seem to be confirming this,
apparent pulse shapes can be simulated and compared to those that we
observe. I look forward to seeing the full results of this study in
the near future. A useful check of this model would be to simulate the
Parkes multibeam survey.

As emphasised by Cordes, even non-detections from pulsar surveys are
interesting results and pulsar hunters at all wavelengths should
continue to write up their negative results (boring though the task
of writing them may be!) for use in likelihood analyses.  A good
example in this regard is the modelling of the gamma-ray pulsar
luminosity function based on OSSE and EGRET results presented by
McLaughlin. This study provides a novel means of constraining a number
of population parameters, such as the initial spin period of pulsars.

Our understanding of poorly-sampled subsets of the
Galactic population, such as the double neutron star systems,
is receiving a significant boost from the work presented by Kalogera.
This population is presently dominated by two objects: the original
binary pulsar B1913+16 and B1534+12. Kalogera and Narayan are presently
addressing the burning question ``how representative are these pulsars
of the underlying sample of objects?'' --- a novel approach to this
is to create a model population (whose underlying parameters are
well understood) and estimate the size of the underlying sample
based on a few objects. It turns out that there is a significant
bias which can be derived from this model, and applied to the
true sample of objects to infer the Galactic population. This is
an important new result and could also be applied to other small samples.

One such application is the population of long-period pulsars. 
We heard from Young that the period of PSR J2144$-$3933,
originally discovered in the Parkes Southern Sky Survey, is 8.5 s ---
three times that previously thought. This is presently the longest
period for a radio pulsar. Young et al.~make the valid point that such
pulsars could be very numerous in the Galaxy since they have very
narrow emission beams and therefore radiate to only a small fraction
of the celestial sphere. Why this pulsar is radiating at all challenges
some current neutron star emission theories and equations-of-state.

Theoretical studies concerning globular cluster pulsar demography and
dynamics are beginning to become popular again. As we heard during
Rasio's talk, the recent flurry of new discoveries in 47~Tucanae and
improved timing solutions are allowing more detailed studies to be
carried out. The radial distribution of pulsars in this cluster out to
a few core radii seems to be consistent with an isothermal sphere.
47~Tucanae is a cluster that has not yet gone through a core collapse
phase and is supporting itself by ``burning'' binaries (thereby
releasing kinetic energy) in the core. Among the questions posed is
why there are so many short-period binaries in 47~Tucanae but
relatively few low-mass X-ray binaries. In addition, there is a
distinct absence of long-period ($\gapp 3$ day) binaries. Presumably,
the latter objects get quickly disrupted during exchange interactions,
which may in turn result in short-period binaries and solitary
millisecond pulsars. Fossil evidence for such interactions may be the
small, but significant, eccentricities now measurable for some of the
binaries in 47~Tucanae.

\section{The Future}

This review has (shamelessly!) focused on radio pulsar astronomy --- a
classic summariser selection bias.  Whilst the current status of this
field is alive and well, giant leaps are anticipated in the future
with the Square Kilometer Array. Funding and radio frequency
interference permitting, there should be no shortage of tasks to be
carried out in the near and long-term future of this field. High
energy astronomy is already embarking on a busy period with the new
generation of satellites, in particular CHANDRA, as we heard is
already producing fantastic results. As already discussed,
gravitational wave astronomy should be increasingly augmenting the
science presented at these meetings in future as new detectors
come on-line. In short, the future of pulsar astronomy looks well set.

\section{And finally...}

The friendly atmosphere which prevailed during the week inevitably
included some lighthearted moments. Three particularly
memorable quotes from the speakers are listed below for posterity:

\bigskip
\noindent
``{\sl The only good pulsar is a dead pulsar!}'' --- G.~Pavlov

\bigskip
\noindent
``{\sl We obtained a good fit to the data with only 18 free parameters.}'' 
--- A.~Somer

\bigskip
\noindent
``{\sl We have convinced ourselves that this will probably work.}'' 
--- J.~Cordes
\bigskip

\noindent
Joking and back-slapping aside, I'm sure that these proceedings will serve
as a useful testament to the events of this meeting and a snap-shot of
the field as it stands. This is particularly important not only
for researchers wanting to catch up on the latest results, but
also for the continual influx of young researchers to
pulsar astronomy, many of whom were present at this meeting.

\bigskip
\noindent
Und {\sl dat}, as they say in Rheinland, war es denn.  I'd like to
close by saying what a pleasure it was to participate in this meeting
--- not only because of all the hot scientific results being
discussed, but also because of the efficient way in which it was
organised and run throughout the week. Whilst many would expect
nothing less from a conference held in Germany, nevertheless the
standard set here will be hard to surpass in future meetings. On
behalf of all of the participants, I'd like to thank all the members
of the Local Organising Committee who put themselves at our disposal
during this week. In particular, hearty thanks go out to Michael
Kramer, Norbert Wex, Gabi Breuer, Ute Runkel and Richard Wielebinski,
all of whom must have breathed a sigh of relief, perhaps even saying
``{\sl nie wieder!}'', when the meeting was over and the last
participant had shuffled out of the building. They can pride
themselves on a job well done and be assured that we are all eagerly
looking forward to another pulsar IAU meeting in Bonn at some point 
early in the next millennium!

\end{document}